# Resonance as a Means of Distance Control in Putting

Robert D. Grober

Department of Applied Physics, Yale University, New Haven, CT 06520

(March 14, 2011)

Introduction

This paper extends an initial study [1] where it was shown the tempo of the putting strokes of accomplished golfers can be explained by assuming these golfers drive their biomechanical system (i.e. golfer and putter) resonantly. In this paper it is proposed that a reason for driving the system resonantly is to simplify distance control. In particular, this paper proposes the distance traveled by the ball, $x_b$, is related to the properties of the putting stroke through the relation $x_b \propto \left(\dfrac{x_s}{\tau_{ds}\langle\langle\alpha\rangle\rangle}\right)^2$, where $x_s$ is the length of the putting stroke, $\tau_{ds}$ is the duration of the downswing, and $\langle\langle\alpha\rangle\rangle$ is the ratio of the average velocity to the final velocity of the club head during the downswing. By driving the system resonantly the golfer can fix $\tau_{ds}$ and $\langle\langle\alpha\rangle\rangle$ to be constant, yielding the single parameter scaling relation $x_b \propto x_s^2$, and thereby simplifying the problem of distance control in putting.

An appendix to this paper describes how golfers can implement, and thereby make use, of the proposed scaling relation.

The Problem: Distance Control in Putting

A goal of this paper is to define properties of a putting stroke which are relevant to controlling how far a putt will roll. Before defining the putting stroke, it is useful to summarize some properties of the ball as it rolls across the green. Suppose the goal is for the ball to roll a distance $x_b$ before coming to rest. The initial speed of the ball $v_{0,b}$ can be calculated by assuming the ball is decelerated by a constant force, $F_b$ [2]. The initial velocity is estimated by using conservation of energy, $\frac{1}{2}mv_{0,b}^2 = F_b\, x_b$, yielding the relation $x_b \propto v_{0,b}^2$. It is assumed $F_b$ is a constant force, independent of the length of the putt, and can therefore be ignored in the scaling relations. Thus, when comparing putts of different lengths within the context of this paper, it is always assumed we are comparing putts on similar putting surfaces, and thus equal $F_b$.

The golfer generates this initial ball speed through collision between the ball and the club head of the putter. While the details of impact between the club head and the ball can be complicated, it can be reasoned that the initial velocity of the ball is proportional to the velocity of the putter head at impact, $v_{0,s} \propto v_{0,b}$, from which the following relation is obtained,

$$x_b \propto v_{0,s}^2. \tag{1}$$

Therefore, a goal of distance control in putting is to deliver the club head to impact at the proper speed, and Eq. (1) suggests the resulting distance traveled by the ball should scale as the square of the speed of the club head at impact.

Characterizing the Motion of the Putting Stroke

Consider the motion of the club head during the downswing. Define $t = 0$ as the point of transition from backswing to downswing, at which the club head velocity is zero. Define $t = \tau_{ds}$ as the moment of impact. Let $x_s$ be the distance from transition to impact, which is the length of the putting stroke.

Assume the acceleration of the club head can be expressed as $a(t) = a_0 \alpha(t)$, where $\alpha(t)$ is a dimensionless function that characterizes the time dependence and is defined such that $\langle \alpha \rangle = \dfrac{1}{\tau_{ds}} \displaystyle\int_0^{\tau_{ds}} dt\, \alpha(t) = 1$. Therefore, $a_0$ is the average acceleration during the downswing. Finally, define the dimensionless parameter $\langle\langle \alpha \rangle\rangle$ to be the second integral of the time dependence of the acceleration,

$$\langle\langle \alpha \rangle\rangle = \frac{1}{\tau_{ds}} \int_0^{\tau_{ds}} dt\, \frac{1}{\tau_{ds}} \int_0^t dt'\, \alpha(t'), \tag{2}$$

The velocity of the club head at impact is

$$v_{0,s} = v(\tau_{ds}) = \int_0^{\tau_{ds}} dt\, a_0\, \alpha(t) = a_0 \tau_{ds} \langle \alpha \rangle = a_0 \tau_{ds}, \tag{3}$$

and the length of the putting stroke is

$$x_s = x(\tau_{ds}) = \int_0^{\tau_{ds}} v(t)\, dt = \int_0^{\tau_{ds}} dt \int_0^t a(t')\, dt' = a_0 \tau_{ds}^2 \langle\langle \alpha \rangle\rangle. \tag{4}$$

A convenient way of understanding $\langle\langle \alpha \rangle\rangle$ is as the ratio of the average velocity of the club head during the downswing to the velocity of the club head at impact,

$$\langle\langle \alpha \rangle\rangle = \frac{1}{\tau_{ds}} \int_0^{\tau_{ds}} dt\, \frac{1}{\tau_{ds}} \int_0^t dt'\, \alpha(t') = \frac{\dfrac{1}{\tau_{ds}} \displaystyle\int_0^{\tau_{ds}} dt \int_0^t dt'\, a_0 \alpha(t')}{a_0 \tau_{ds}} = \frac{\langle v \rangle}{v_{0,s}}.$$

Thus, the sooner the club head settles to its final velocity, the closer $\langle\langle\alpha\rangle\rangle$ will be to unity.

Combining (3) and (4) yields

$$v_{0,s} = \frac{x_s}{\tau_{ds}\langle\langle\alpha\rangle\rangle}. \tag{5}$$

Combining (5) and (1) results in the expression

$$x_b \propto \left(\frac{x_s}{\tau_{ds}\langle\langle\alpha\rangle\rangle}\right)^2 \tag{6}$$

Thus, for the purposes of this paper, the motion of the club head during the putting stroke is characterized by three parameters: 1) the length of the putting stroke, 2) the duration of the downswing, and 3) the ratio of the average velocity to the final velocity of the club head during the downswing. Changing any of these parameters will change the distance that the ball rolls according to the scaling relation defined in Eq. (6). That $x_b$ scales as the square of the three independent putting stroke parameters $x_s$, $\tau_{ds}$, and $\langle\langle\alpha\rangle\rangle$ explains why distance control can be very difficult: there are three adjustable parameters and $x_b$ depends non-linearly on all three of them. Clearly, a key to distance control is to simplify this situation.

The Uniform Acceleration Putting Stroke

As a simple example of the use of Eq (5) and (6), consider the case of a putting stroke where the motion of the putter during the downswing is characterized by a uniform acceleration, $a_0$, for the duration of the downswing, $\tau_{ds}$. Under these conditions, the speed of the putter at impact is $v_{0,p} = a_0 \tau_{ds}$ and the length of the putting stroke is $x_s = 0.5 \, a_0 \tau_{ds}^2$. Eliminating the acceleration in these expressions yields

$$v_{0,s} = \frac{x_s}{0.5 \, \tau_{ds}}. \tag{7}$$

Note that for the case of uniform acceleration, $\langle\langle a \rangle\rangle = \frac{1}{\tau_{ds}} \int_0^{\tau_{ds}} dt \, \frac{1}{\tau_{ds}} \int_0^t dt' = 0.5$, and thus (7) is consistent with (5). Finally, Eq. (6) simplifies to

$$x_b \propto \left( \frac{x_s}{0.5 \, \tau_{ds}} \right)^2 \tag{8}$$

Within the context of this model, the distance the ball travels is dependent on both the length of the putting stroke and the duration of the downswing, which is a function of the magnitude of the acceleration. There are an infinite number of possible putting strategies which yield the desired result: short putting strokes require large acceleration while long putting strokes require small acceleration. The problem confronting the golfer is that the abundance of choices makes regularizing the choice of strategy difficult.

The Constant Acceleration Putting Strategy for Distance Control

As an example of one such possible strategy, consider a golfer who always uses the same constant acceleration, $a_0$, for every putting stroke regardless of the length of the putt. To be clear, this is just an example and likely not an advisable strategy. For a putting stroke of length $x_s$, the velocity at impact is $v_{0,s}^2 = 2 a_0 x_s$. From Eq (1), the distance traveled by the ball is proportional to the length of the putting stroke, $x_p \propto x_s$, which seems intuitive. It follows from $x_s = 0.5 \, a_0 \tau_{d,s}^2$ that the duration of the downswing using this putting strategy is proportional to the square root of the length of the putting stroke, $\tau_{ds} \propto \sqrt{x_s}$, which also seems like a reasonable outcome. However, a drawback of this strategy is that there is no physical means of assuring constant $a_0$ from putt-to-putt and from day-to-day.

In contrast to this example, the resonant putting stroke uses mechanical resonance as a timing reference, allowing the golfer to reliably set $\tau_{ds}$ to be the same value from putt-to-putt and from day-to-day. As is described below, this resonant putting stroke also forms the basis of a repeatable strategy for distance control.

The Resonant Putting Stroke

In [1] it was proposed that a reliable way to fix $\tau_{ds}$ is to drive the putting stroke at a mechanical resonance of the system, thus using the physical system as the timing reference. As is shown below, an additional property of this type of stroke is that $\langle\langle \alpha \rangle\rangle$ is also the same from putt to putt. According to Eq. (6), if the golfer is able to develop a putting technique such that $\tau_{ds}$ does not vary from putt-to-putt, and $\langle\langle \alpha \rangle\rangle$ is the same from putt-to-putt, the scaling relation of Eq. (6) simplifies to $x_b \propto x_s^2$.

Summarizing the analysis in [1], the equation of motion $\ddot{x} + \omega_0^2 x = \frac{f(t)}{m}$ is solved subject to the initial conditions $x(0) = 0$ and $\dot{x}(0) = 0$. The applied force is given as

$$f(t) = \begin{cases} -f_0 \sin 2\omega_0 t & 0 < \omega_0 t < \pi \\ 0 & \text{otherwise} \end{cases} \qquad (9)$$

and the resulting solution is $x(0 < \omega_0 t < \pi) = -\dfrac{v_{0,s}}{4\omega_0}(2\sin \omega_0 t - \sin 2\omega_0 t)$, where $v_0$ is the

velocity of the club at impact. The transition from backswing to downswing occurs at

$t = \tau_{bs} = \dfrac{2\pi}{3\omega_0}$ and impact occurs when putter returns to the origin at $t = \tau_{bs} + \tau_{ds} = \dfrac{\pi}{\omega_0}$.

Thus, the duration of the downswing is $\tau_{ds} = \dfrac{\pi}{3\omega_0}$ and the tempo ratio $\dfrac{\tau_{bs}}{\tau_{ds}} = 2$.

It is useful to write the acceleration in the form

$$a(0 < \omega_0 t < \pi) = \dfrac{v_{0,s}}{\tau_{ds}}\dfrac{\pi}{6}(\sin \omega_0 t - 2\sin 2\omega_0 t) = \dfrac{v_{0,s}}{\tau_{ds}}\alpha(t) \qquad (10)$$

where we identify $\dfrac{v_{0,s}}{\tau_{ds}}$ as the average acceleration during the downswing and

$$\alpha(t) = \dfrac{\pi}{6}(\sin \omega_0 t - 2\sin 2\omega_0 t) \qquad (11)$$

which can be recast as

$$\alpha(t) = \dfrac{\pi}{6}\left(\sin \dfrac{\pi t}{3\tau_{ds}} - 2\sin \dfrac{2\pi t}{3\tau_{ds}}\right). \qquad (12)$$

It is straightforward to verify

$$\langle \alpha \rangle = \dfrac{1}{\tau_{ds}}\int_{2\tau_{ds}}^{3\tau_{ds}} dt\, \alpha(t) = 1 \qquad (13)$$

and

$$\langle\langle \alpha \rangle\rangle = \dfrac{1}{\tau_{ds}}\int_{2\tau_{ds}}^{3\tau_{ds}} dt\, \dfrac{1}{\tau_{ds}}\int_{2\tau_{ds}}^{t} \alpha(t')\,dt' = \dfrac{9\sqrt{3}}{8\pi} \approx 0.62. \qquad (14)$$

According to Eq. (5), the relation between the length of the putting stroke, the velocity of

the club head at impact, and the duration of the downswing is

$$v_{0,s} = \frac{x_s}{0.62\,\tau_{ds}}, \tag{15}$$

not very different than that for the case of constant acceleration, Eq. (7).

Finally, Eq (6) for the case of the resonant putting stroke becomes

$$x_b \propto \left(\frac{1}{0.62\,\tau_{ds}}\right)^2 x_s^2, \tag{16}$$

where the term in parenthesis is constant from putt to putt. Thus, the distance the ball travels is now only a function of the length of the putting stroke, $x_b \propto x_s^2$, which simplifies the problem of distance control to a single parameter scaling relation. Note that this implies a putting stroke which is more compact than the constant acceleration example cited above. Additionally, the foundation of this strategy is the use of mechanical resonance to fix $\tau_{ds}$, which enables consistency from putt-to-putt and from day-to-day.

It is interesting to note that in addition to fixed $\tau_{ds}$, it has been observed that many putting strokes exhibit large regions of the downswing during which the acceleration is uniform [3]. This type of putting stroke can be thought of as a small perturbation to the downswing of the resonant putting stroke. Because $\tau_{ds}$ is a constant independent of $x_s$, this acceleration presumably increases linearly with $x_s$, which preserves the relation $x_b \propto x_s^2$. However, for these putting strokes $\langle\langle\alpha\rangle\rangle$ will likely fall in the range $0.5 \le \langle\langle\alpha\rangle\rangle \le 0.62$, i.e. the downswing is some mixture between pure resonance and uniform acceleration. This narrow range of value reflects the fact that the difference between these putting strokes is relatively small.

The Resonant Putting Stragegy for Distance Control

The above analysis suggests that when employing a resonant putting stroke, the ball will roll a distance that scales as the square of the length of the putting stroke. Table 1 provides a simple guideline for understanding this scaling relation within the context of hitting real putts. The table lists the distance traveled by the ball, $x_b$, as a function of the length of the putting stroke $x_s$ assuming the relationship $x_b \propto x_s^2$. The values of $x_s$ extend from 4" to 24" in increments of 2". The values of $x_b$ are listed in arbitrary units so that one can adapt them to any putting condition (i.e. the Stimp speed of a particular green, uphill or downhill surfaces, individual putting tempo, etc.).

| Length of Putting Stroke, $x_s$ (inches) | Distance Traveled by Ball, $x_b$ (arb. unit) |
|---|---|
| 4" | 1 |
| 6" | 2.25 |
| 8" | 4 |
| 10" | 6.25 |
| 12" | 9 |
| 14" | 12.25 |
| 16" | 16 |
| 18" | 20.25 |
| 20" | 25 |
| 22" | 30.25 |
| 24" | 36 |

Table 1: The distance traveled by the ball, in arbitrary units, as a function of the length of the putting stroke in inches.

As an example of its use, if an 8 inch backswing consistently yields an 8 foot putt, then a 12 inch backswing should yield a putt that is longer by the ratio 9/4 (i.e. an 18 foot foot putt) and a 16 inch backswing should yield a putt that is longer by the ratio 16/4 (i.e. a 32 foot putt).

Note that Table 1 is relevant to putts of different lengths on a flat surface. It does not matter whether the surface is level, uphill, or downhill, so long as the surface is flat.

As was described above, the numbers in this table work only in the limit that $\tau_{ds}$ and $\langle\langle\alpha\rangle\rangle$ are independent of $x_s$. Conversely, if your putting is consistent with the numbers in the chart, then you can be confident you are putting such that $\tau_{ds}$ and $\langle\langle\alpha\rangle\rangle$ are independent of $x_s$.

The appendix to this paper describes how one can learn what it is to putt with a resonant putting stroke.

Testable Hypotheses

This analysis suggest several useful measurements of the putting stroke. The tempo of the putting stroke is often characterized as the ratio of the duration of the backswing $\tau_{bs}$ to the duration of the downswing $\tau_{ds}$, and it is well documented that this ratio is of order two for the putting stroke. This analysis suggests it is also useful to measure the length of the putting stroke $x_s$, the velocity of the club head at impact $v_{0,s}$, the ratio of the average velocity to the final velocity of the club head during the downswing $\langle\langle\alpha\rangle\rangle$, and the distance the putt rolls, $x_b$. Subsequent analysis should test the following three hypotheses:

1) The relation $x_b \propto \left(\dfrac{x_s}{\tau_{ds} \langle\langle\alpha\rangle\rangle}\right)^2$ should hold generically for any putting stroke.

2) $\langle\langle\alpha\rangle\rangle$ and $\tau_{ds}$ should be relatively constant, independent of $x_b$, in the most accomplished of golfers, as this is an optimal strategy for distance control.

3) $\langle\langle\alpha\rangle\rangle$ should be of order 0.5 for a putting stroke characterized by uniform acceleration during the downswing and of order 0.62 for a purely resonant putting stroke. In practice, it is likely $\langle\langle\alpha\rangle\rangle$ will fall in a range between these two extremes.

Conclusion

In this paper it is reasoned that the distance traveled by the ball during a putt $x_b$, depends on the length of a putting stroke $x_s$, the duration of the downswing $\tau_{ds}$, and the ratio of the average velocity to the final velocity of the club head during the downswing $\langle\langle\alpha\rangle\rangle$ through the relation $x_b \propto \left(\dfrac{x_s}{\tau_{ds} \langle\langle\alpha\rangle\rangle}\right)^2$. It has previously been observed that capable golfers use mechanical resonance to putt with tempo which is independent of the length of the putt, effectively setting $\tau_{ds}$ equal to a constant. In this paper we show that a resonant putting stroke also fixes $\langle\langle\alpha\rangle\rangle$ to be a constant of order 0.62. This simplifies the scaling relation, with $x_b$ depending only on the square of the length of the putting stroke, $x_b \propto x_s^2$, thereby simplifying the problem of distance control in putting.

Appendix I: Towards A Resonant and Balanced Putting Stroke

Between this paper and its predecessor [1] two principles guiding the dynamics of the putting stroke are obtained. The first principle is that forces are applied resonantly with the biomechanical system, which consists of the combination of the golfer and the putter. Ref [1] reported the observation that resonance explains many of the quantitative details of putting tempo, including a tempo that is independent of the length of the putt and a putter speed that is constant as the club head approaches impact. In this paper it is reasoned that an additional rational for putting resonantly is to simplify distance control.

The second principle is that the forces should be balanced, i.e. that the backward going and forward going forces should be of the same magnitude. It was shown in [1] that this balancing within the context of a resonant stroke yields the observed 2:1 ratio of the duration of the backswing to the duration of the downswing in the putting stroke. It was also discussed in [1] that a rational for doing this is to minimize the variability in putter speed at impact due to random errors in the application of force.

The following paragraphs describe how one might develop a putting stroke which is both resonant and balanced.

A Resonant Putting Stroke

For most golfers, the primary challenge will be to figure out what it is to swing resonantly. So, lets start by describing resonance within the context of a simple demonstration that can be done by anybody with a golf club. Hold a club, or any reasonably sized stick, with one hand at the very end of the grip by pinching the grip between your thumb and middle finger. Let it hang down with your arm extended away from you body.  (Read it again, as this isn't intended to be a golf swing.)  Now gently let the club swing back and forth, just like the pendulum of a grandfather clock, without moving your body or arm; just with a gentle, subtle, motion of your hand.  Most people will instinctively swing it back and forth at a tempo which requires very little applied force.  See how lightly you can pinch the grip and still get the putter to swing back and forth.

You are swinging it resonantly when you find the tempo at which you have to apply the least amount of force to keep it swinging back and forth.  For most golf clubs, it will take slightly longer than one second for a complete oscillation.  Experiment a little by swinging the club back and forth at much slower speeds and then ridiculously high speeds.  Note that if you swing it back and forth such that a complete cycle takes five seconds, you have to hold on with your entire hand to supply sufficient force.  Thus, if you drive the pendulum at a pace which is too slow, it requires more force to move the club through a comparable distance than if you drive it resonantly.  Similarly, if you swing the club back and forth much too fast you will note that the club head doesn't move very far no matter how hard you drive it.

Swinging the club resonantly means swinging it back and forth at a tempo that uses the least amount of force to obtain the intended amplitude of motion.  Note that once you are resonant, you can maintain perfectly reproducible tempo pretty much without paying attention; using as little force as is possible seems to be instinctive and the resonance imposes the reproducible tempo.  Also note that when you swing the club resonantly, it takes the same amount of time for a small swing as it does for a big swing.  This is a fundamental property of a system driven at resonance.  This is why a grandfather clock keeps time. This is why the tempo of the putting stroke of accomplished golfers is very consistent, independent of the length of the putt.

You find can find the resonance in your putting stroke the same way. Take your putting practice stroke by continuously making putter strokes; back and forth, and back and forth; continuously; over and over again; just like you did for the pendulum. Experiment by speeding up and slowing down the tempo until you find the pace at which you use the least amount of force to make the stroke, i.e. <u>*make the stroke feel as effortless as is possible*</u>. Note that the biomechanical system can be slightly more complicated that a simple pendulum. For instance, you can make the system stiffer by tightening your muscles, thereby increasing the natural resonant frequency.

For most golfers, the resonant tempo is faster than what they expect. So, experiment a bit, changing the tempo dramatically. Speed it up way too fast and slow it down to ridiculously slow speeds. If you experiment for a while, you will eventually find that tempo which feels effortless, and that is likely a resonantly tempo.

Once you find the resonant tempo, try making a real putting stroke using exactly this tempo. Start with the club at the position of the ball and then make the backswing, downswing, and follow-thru using the same tempo you found earlier. If after a couple of putts you feel you have lost the feel for the tempo, go back to swinging the club continuously back and forth until you find the tempo again.

<u>A Balanced Putting Stroke</u>

If the force you use to push back in the backswing is the same magnitude as the force you use to push down in the downswing, and if you are working at a resonant tempo, then the ratio of the duration of the backswing to the duration of the downswing will be about 2:1. This 2:1 ratio is what has been observed for accomplished golfers [1]. You can get a sense of this 2:1 ratio with a simple counting experiment. If you count '1-2-3-4' rhythmically, evenly spacing the distance between each number, such that '1' happens at the start of the backswing, '3', happens at the transition from backswing to downswing, and '4' happens at impact, then the ratio of backswing to downswing time will be about 2:1. It doesn't have to be perfect; getting it approximately correct is close enough. The key is to balance the backward and forward going forces.

Now, try to swing with this same tempo for both short putting strokes and for long putting strokes. Use the same '1-2-3-4' counting technique to gauge the 2:1 ratio and the tempo. Remember that the 2:1 ratio happens when the backward-going and forward-going forces are balanced. If you feel like you have lost the tempo, go back to swinging back and forth over and over again like a pendulum until you find the tempo again.

Summary

The goal is to develop a putting stroke which is both resonant and balanced. Resonance serves many purposes, one of which is to make the duration of the putting stroke independent of the length of the putt. This considerably simplifies the strategy for distance control. Balancing the backward and forward going forces serves to minimize variation in club head speed at impact due to small errors in the application of force.

An excellent test of whether this is what you have done is to test your stroke against the guidelines in Table 1. If you find that when you double the length of your backswing the putt goes four times further; and if you triple the length of your backswing the putt goes nine times further; and if your quadruple the length of your backswing the ball goes sixteen times further; and if this all happens reproducibly; then you have developed a putting stroke optimized for distance control.